\documentstyle[preprint,prd,aps,psfig,epsfig]{revtex}
\tighten
\let\jnfont=\rm
\def\NPB#1,{{\jnfont Nucl.\ Phys.\ }{\bf B#1},}
\def\PLB#1,{{\jnfont Phys.\ Lett.\ B }{\bf #1},}
\def\PRD#1,{{\jnfont Phys.\ Rev.\ D }{\bf #1},}
\def\PRL#1,{{\jnfont Phys.\ Rev.\ Lett.\ }{\bf #1},}
\def\ZPC#1,{{\jnfont Z.~Phys.\ C }{\bf #1},}
\newcommand{\gsim}{\mathrel{\lower4pt\hbox{$\sim$}}
\hskip-12.5pt\raise1.6pt\hbox{$>$}\;}
\newcommand{\lsim}{\mathrel{\lower4pt\hbox{$\sim$}}
\hskip-12.5pt\raise1.6pt\hbox{$<$}\;}
\begin{document}
\preprint{\parbox{2.0in}{\noindent TU-580\\ NUHEP-TH-99-26\\ 
                                            AMES-HET-99-10 \\ }}
\title{\ \\[15mm] Implications of LEP/SLD Data for New Physics 
        in $Zb\bar b$ Couplings}
\vspace{1cm}
\author{\ \\[2mm] Robert J. Oakes$^a$, Jin Min Yang$^{b,c}$,
                                        Bing-Lin Young$^{d}$ }
\address{\it \ \\[2mm] 
$^a$ Department of Physics and Astronomy, Northwestern University,
         Evanston, IL 60208, USA\\ 
$^b$ Department of Physics, Tohoku University,
          Aoba-ku, Sendai 980-8578, Japan\\
$^c$ Institute of Theoretical Physics, Academia Sinica,
          Beijing 100080, China\\
$^d$ Department of Physics and Astronomy, Iowa State University,
     Ames, Iowa 50011, USA}
\maketitle
\vspace{2cm}

\begin{abstract}
The combined LEP/SLD data on the $b\bar b$ forward-backward asymmetry
from the Z-pole measurements may imply the presence of new physics 
in the $Zb\bar b$ couplings.  In general, the effect of new physics can 
be parameterized by $SU_C(3)\times SU_L(2)\times U_Y(1)$ invariant higher 
dimensional operators. By fitting the recently announced LEP/SLD data on
$A_b$ and $R_b$, the size of the 
coupling strengths of these operators can be determined.  We also found that 
the new physics operators can be divided into two types, depending on
their Higgs field content. The ones involving the Higgs have very mild effects 
at higher energy colliders, while the other type which do not contain the 
Higgs field can show significantly large effects on $b\bar b$ production at 
LEP II, $t\bar t$ production at the NLC and single top production at the 
Tevatron. The preliminary data from 
the LEP II measurements disfavors the second type of operators.
\end{abstract}
\newpage

\section{Introduction}
\label{sec1}
The continuing agreement between experimental data and theoretical predictions 
on almost all the electroweak variables has further solidified the standard 
model (SM). However, there are signs of perhaps something non-standard
from the investigation of non-electroweak variables and the indication of 
possibilities of new physics from electroweak variables themselves as well.  
On the one hand, the 
accumulative and, especially, the recent data on neutrino oscillations have 
strongly suggested the fact of finite neutrino masses \cite{peccei} and,
therefore, the SM has to be modified.  On the other hand, the $Z$-pole 
measurements at LEP/SLD on the $b\bar b$ forward-backward asymmetry gives a 
value of 
$A_b=\left (g^2_L(b) - g^2_R(b)\right )/\left ( g^2_L(b) + g^2_R(b)\right )$ 
which deviates from the SM prediction by $2.7 \sigma$ \cite{data}.   
If this $A_b$ anomaly is not a statistical or systematic effect, 
it signals the presence of new physics in association with the $Zb\bar b$
coupling.   Meanwhile, the experimental value of the related quantity 
$R_b= \Gamma(Z\to b\bar b)/\Gamma(Z\to {\rm hadrons})$,
after showing a deviation from the Standard Model (SM) value for a few years, 
now agrees well with the SM prediction \cite{data}.   
To explain the current experimental
values of both $A_b$ and $R_b$, the required new physics contribution has to 
shift the left- and right-handed  $Zb\bar b$ couplings by  $\sim -1\%$ and  
$\sim +30\%$, respectively\cite{mar,chang}.  Recently efforts have been made 
in exploring the $A_b$ anomaly, either in explaining the effect in specified 
models \cite{chang,A_b} or in examining its implications for low energy 
decay processes \cite{chanowitz}. The popular low energy SUSY
and the models in which the third generation feels a different
gauge dynamics from the usual weak interaction
cannot yield such large anomalous  $Zb_R\bar b_R$ coupling,
as showed in \cite{Zbb_SUSY} and \cite{yuan}, respectively.   
These studies are focused on the theme 
that has been emerged since the discovery of the heavy top quark, 
that new physics is most likely 
related to the gauge symmetry breaking sector and, therefore, may affect 
the interactions of the third family quarks most significantly.

From the overwhelming success of the SM at the electroweak scale, one can
conclude that the underlying theory of the new physics can openly 
manifest only at a higher energy scale.  One can envisage in the scheme 
of a larger symmetry that
encompasses the new physics, after integrating out the heavy degrees of 
freedom which lie outside the SM spectrum, higher dimensional terms will be 
present at energies not too far above the electroweak scale and the induced
effective terms should preserve the basic SM structure. 
Then the new physics effects on the $Zb\bar b$ couplings can be parameterized 
by a set of higher dimensional operators \cite{dim6,dim6_b} which are  
$SU_C(3)\times SU_L(2)\times U_Y(1)$ symmetric before the electroweak 
symmetry breaking becomes explicit. 
From the above argument we see that before the electroweak symmetry breaking, 
the $Zb\bar b$ part of the effective Lagrangian can take the general form 
\begin{equation}\label{eq1}
{\cal L}^{eff}_{Zb\bar b}={\cal L}^{SM}_{Zb\bar b}
                         +\frac{1}{\Lambda^2}\sum_i C_i O_i
                         +O(\frac{1}{\Lambda^4})
\end{equation}
where ${\cal L}^{SM}_{Zb\bar b}$ is the SM part, $\Lambda$ is the new physics 
scale, $O_i$ are dimension-six SM gauge invariant operators,
and $C_i$ are constants which represent the coupling strengths of $O_i$.
We have assumed that operators of higher dimensions than 6 are suppressed
by powers of $1/\Lambda^2$, so Eq. (\ref{eq1}) is a quite general 
parameterization of new physics.  

Since there are many possible higher dimensional operators $O_i$, it is 
important to use various experimental data to constrain the operator form in 
order to narrow down the directions of the underlying theory.
Because the operator $O_i$ can contribute to several vertices after  
$SU_L(2)\times U_Y(1)$ symmetry breaking, these operators can contribute
to and will show correlated effects on several different physical observables.
It will become clear later that the new physics 
effects required to explain $A_b$ will also show up in other observables,
such as $b\bar b$ production at LEP II, $t\bar t$ production at NLC,
and $t\bar b$ production at the Tevatron. These effects help to distinguish
the possible operator forms required. 

This article is organized as follows. In Sec. \ref{sec2} we list the
contributing dimension-6 operators and derive their induced vertices.
In Sec. \ref{sec3} we constrain the coupling strengths of the operators 
using the LEP/SLD  Z-pole $Zb\bar{b}$ data.  In Sec. \ref{sec4} we investigate 
the correlated effects of these operators at LEP II, the NLC, and the 
Tevatron. In Sec. \ref{sec5} we present our discussions and the conclusion.   
 
\section{The operators and their induced vertices}
\label{sec2}

Assuming CP conservation and ignoring those operators which only lead 
to anomalous dipole-moment couplings which are suppressed, 
we have two dimension-six gauge invariant operators which give rise 
to anomalous right-handed $Zb\bar b$ coupling\cite{dim6},
\begin{eqnarray}\label{bB}
O_{bB}&=&\left [\bar b_R \gamma^{\mu} D^{\nu}b_R
         +\overline{D^{\nu}b_R} \gamma^{\mu}b_R\right ]
          B_{\mu\nu},\\ \label{phib}
O_{\Phi b}&=&i\left [\Phi^{\dagger}D_{\mu}\Phi
         -(D_{\mu}\Phi)^{\dagger}\Phi\right ]\bar b_R \gamma^{\mu}b_R,
\end{eqnarray}
and four operators affecting the left-handed $Zb_L\bar b_L$ coupling, 
\begin{eqnarray}\label{qW}
O_{qW}&=&\left [\bar q_L \gamma^{\mu}{\sigma^I\over 2} D^{\nu}q_L
         +\overline{D^{\nu}q_L} \gamma^{\mu}{\sigma^I\over 2}q_L\right ]
          W^I_{\mu\nu},\\
O_{qB}&=&\left [\bar q_L \gamma^{\mu} D^{\nu}q_L
         +\overline{D^{\nu}q_L} \gamma^{\mu} q_L\right ]
          B_{\mu\nu},\\ \label{phiq1}
O_{\Phi q}^{(1)}&=&i\left [\Phi^{\dagger}D_{\mu}\Phi
      -(D_{\mu}\Phi)^{\dagger}\Phi\right ]\bar q_L \gamma^{\mu}q_L,\\ 
 \label{phiq3}
O_{\Phi q}^{(3)}&=&i\left [\Phi^{\dagger}{\sigma^I \over 2} D_{\mu}\Phi
        -(D_{\mu}\Phi)^{\dagger}{\sigma^I\over 2}\Phi\right ]
        \bar q_L \gamma^{\mu}{\sigma^I \over 2} q_L.
\end{eqnarray}
At the order of $C_i/\Lambda^2$, the contributions to $R_b$ and 
$A_b$ from those operators that only lead to anomalous dipole-moment 
couplings for $Zb\bar b$ are suppressed by a factor $m_b/m_Z$ relative 
to the contributions of the operators we are considering.
We have used the conventional notation:  
$q_L=(t_L, b_L)$ is $SU_L(2)$ doublet of the third family of quarks; 
$\Phi$ the Higgs boson doublet; $B_{\mu}$ the $U_Y(1)$ gauge field and 
$B_{\mu\nu}=\partial_{\mu}B_{\nu} -\partial_{\nu}B_{\mu}$; 
$W^I_{\mu}$ ($I=1,2,3$) are the $SU_L(2)$ gauge fields and 
$W^I_{\mu\nu}=\partial_{\mu}W^I_{\nu}-\partial_{\nu}W^I_{\mu}        
+g_2\epsilon_{IJK}W^J_{\mu}W^K_{\nu}$, and $\sigma^I$ are the Pauli matrices. 
  
After the electroweak symmetry breaking each operator can give rise to a set 
of vertices which may affect more than one electroweak observable involving 
the bottom quark and lead to correlated effects among these observables. 
The two 
right-handed operators, $O_{bB}$ and $O_{\Phi b}$ in Eqs. (\ref{bB}) and 
(\ref{phib}), will give rise to anomalous $Z b_R\bar b_R$ and 
$\gamma b_R\bar b_R$ couplings. The four left-handed operators given in 
Eqs. (\ref{qW})-(\ref{phiq3}) also involve the top quark and therefore will 
contribute to both bottom quark and top quark anomalous couplings.  The 
correlated effects of these operators at the various colliders are depicted 
in Fig.1.

Denoting the $V^0 q\bar q$ $(V^0=Z,\gamma; q=t,b)$ and $Wt\bar b$ vertices by
\begin{eqnarray}\label{Vqq}
 \Gamma^{\mu}_{V^0 q\bar q}&=&-ieg^V\gamma^{\mu}
\left [ P_L (g^V_L+\delta g^V_L) +P_R (g^V_R+\delta g^V_R)\right ],\\
\label{Wtb}
\Gamma^{\mu}_{Wt\bar b}&=&-i\frac{g}{\sqrt 2}
  \gamma^{\mu}P_L(1+\delta g^W_L),
\end{eqnarray}
the SM couplings are  $g^V_L$, $g^V_R$, and  $g^W_L$, while
the anomalous couplings are denoted by $\delta g^V_L$, $\delta g^V_R$ 
and $\delta g^W_L$.  Here $P_{L,R}=(1\mp \gamma_5)/2$,
$g^{\gamma}=1$, $g^{Z}=1/(4s_Wc_W)$ with $s_W\equiv \sin\theta_W$ and 
$c_W\equiv \cos\theta_W$.
The SM couplings are given by $g^{\gamma}_L(q)=g^{\gamma}_R(q)=e_q$, 
$g^Z_L(q)=4I^3_q-4s_W^2e_q$ and $g^Z_R(q)=-4s_W^2e_q$ with $e_q$ being 
the electric charge of the quark in units of $e$ and $I^3_q=\pm 1/2$, the 
weak isospin component.  The new physics contributions to $\delta g^V_R$ 
from the right-handed operators are given in Table \ref{tab1}, and those to 
$\delta g^V_L$ and $\delta g^W_L$  from the left-handed operators in Table 
\ref{tab2}.  The three operators, $O_{bB}$, $O_{qW}$ and $O_{qB}$, induce 
momentum dependent anomalous couplings, and, therefore, their effects 
will generally be enhanced in higher energy processes. 

It should be pointed out that there are no contributions from anomalous $t_R$ 
couplings in the present considerations. The reason is that the operators
which contribute to the $t_R$ coupling are of the form of the operators
given in Eqs.(\ref{bB}) and (\ref{phib}), $b_R$ replaced by $t_R$, and  
do not involve the bottom quark;  we therefore ignore all these operators. 

\section{Constraints from the $Z$-pole data}
\label{sec3}

The non-standard contributions to $R_b$ and $A_b$ at  the  $Z$-pole 
from the anomalous couplings in Eq. (\ref{Vqq}) can be written as
\begin{eqnarray}\label{Rb}
\delta R_b&=&2R_b^{SM}(1-R_b^{SM})
  \left[  \frac{g_L^Z(b) \delta g_L^Z(b)+g_R^Z(b) \delta g_R^Z(b)}
      {\left( g_L^Z(b)\right )^2+\left( g_R^Z(b)\right )^2}\right ],\\ 
\label{A_b}
\delta A_b&=&2A_b^{SM}\left[ 
   \frac{g_L^Z(b) \delta g_L^Z(b)-g_R^Z(b) \delta g_R^Z(b)}
  {\left( g_L^Z(b)\right )^2-\left( g_R^Z(b)\right )^2}
   -\frac{g_L^Z(b) \delta g_L^Z(b)+g_R^Z(b) \delta g_R^Z(b)}
      {\left( g_L^Z(b)\right )^2+\left( g_R^Z(b)\right )^2} \right ]. 
\end{eqnarray}
Here we only keep the lowest order effects of new physics; i.e.,
${\cal O}(C_i/\Lambda^2)$, which is the interference of new
physics terms with the SM contributions, and neglect the interference terms proportional 
to $m_b/m_Z$. The SM values of $R_b^{SM}$ and $A_b^{SM}$ are taken 
to be  \cite{mar}
\begin{eqnarray}\label{data1}
R_b^{\rm SM}=0.2158\pm 0.0002, ~A_b^{\rm SM}=0.9347\pm 0.0001,
\end{eqnarray}
which are the predictions in the SM including radiative corrections.    
The anomalous couplings $\delta g_L^Z(b)$ and $\delta g_R^Z(b)$
in Eqs.(\ref{Rb}) and (\ref{A_b}) are obtained from Tables 1 and 2 
with the $Z$ boson being on mass-shell ($k^2=m_Z^2$). 

The experimental values for $A_b$ and $R_b$ reported in \cite{data} are
\begin{eqnarray}\label{data2}
& & R_b^{\rm exp}({\rm LEP+SLD})=0.21642\pm 0.00073,\\ \label{data3}
& & A_b^{\rm exp}({\rm LEP})=0.881\pm 0.020, ~
    A_b^{\rm exp}({\rm SLD})=0.905\pm 0.026,
\end{eqnarray}
where the LEP value of $A_b$ is obtained from the 
measured quantities $A_{\rm FB}(b)=\frac{3}{4}A_eA_b$ using
$A_e=0.1496\pm 0.0016$, the combined average of $A_e$ from LEP and SLD.
The combined value of $A_b$ from LEP and SLD is then given by
$A_b^{\rm exp}({\rm LEP+SLD})=0.8902\pm 0.0158$, 
which is $2.8\sigma$ below the SM prediction.
To fit the data on both $R_b$ and $A_b$ the  $Zb\bar b$ couplings
are required to be \cite{chang} $g_L^Z(b)/4=-0.4163\pm 0.0020$
and $g_R^Z(b)/4=0.0996\pm 0.0076$. Comparing with the SM values
\cite{field} $g_L^Z(b)/4=-0.4208$ and $g_R^Z(b)/4=0.0774$, 
obtained by including radiative corrections and taking $m_t=174$ GeV 
and $m_H=100$ GeV, 
we find that new physics effects are needed in both $Zb_R\bar b_R$ and 
$Zb_L\bar b_L$, which indicates both right- and 
left-handed higher dimension operators are necessary.
If we use only one right- and one left-handed operator at a time to 
obtain the required modifications to the $Zb_R\bar b_R$ and
$Zb_L\bar b_L$ couplings, we obtain the ranges of strengths of 
the operators at the $1\sigma$ ($2\sigma$) level:    
\begin{eqnarray} \label{limit1}
    2.72~(1.30)\lsim\frac{\vert C_{bB}\vert}{(\Lambda/{\rm TeV})^2} \lsim 
    5.54~(6.96),
~~ 0.48~(0.23)\lsim \frac{\vert C_{\Phi b}\vert}{(\Lambda/{\rm TeV})^2}\lsim  
   0.99~(1.24),                                             \\ \label{limit2}
   0.51~(0.10) \lsim\frac{\vert C_{qW}\vert}{(\Lambda/{\rm TeV})^2}\lsim 
   1.33~(1.73),
~~ 0.47~(0.09)\lsim\frac{\vert C_{qB}\vert}{(\Lambda/{\rm TeV})^2}\lsim 
   1.21~(1.58),                                             \\ \label{limit3}
   0.08~(0.02 )\lsim\frac{\vert C_{\Phi q}^{(1)}\vert}{(\Lambda/{\rm TeV})^2}
\lsim ~0.22~(0.28),~~
0.08~(0.02)\lsim\frac{\vert C_{\Phi q}^{(3)}\vert}{(\Lambda/{\rm TeV})^2}
           \lsim ~0.22~(0.28).
\end{eqnarray}

Apart from the limits that can be obtained from the electroweak variables
the strengths of the operators can also be constrained by the partial wave
unitarity condition of the appropriate 2-to-2 scattering processes
$b\bar b\leftrightarrow b\bar b$, $b\bar b\leftrightarrow t\bar t$
and $t\bar t\leftrightarrow t\bar t$, which 
involve the helicity channels $b_+\bar b_-$,
$b_-\bar b_+$, $t_+\bar t_-$ and $t_-\bar t_+$, with $+$ and $-$ 
denoting positive and negative helicities \cite{unitary}.  
For the three operators, $O_{bB}$, $O_{qW}$, and $O_{qB}$,
which give rise to momentum dependent couplings, the unitarity constraints
are found to be significant \cite{unitary}, and are given by
\begin{eqnarray}
& &  \frac{\vert C_{bB}\vert}{\Lambda^2}<\frac{\sqrt{8\pi}}{s},\\
& &  \frac{\vert C_{qW}\vert}{\Lambda^2}<\frac{\sqrt{4\pi}}{s},
\end{eqnarray}
and
\begin{eqnarray}
& &  \frac{\vert C_{qB}\vert}{\Lambda^2}<\frac{\sqrt{4\pi}}{s}.
\end{eqnarray}
Here $s$ is the center-of-mass energy squared for the relevant process.
Requiring the unitarity condition to be satisfied for the processes 
with  center-of-mass energy up to the new physics scale; i.e., 
$\sqrt s\approx\Lambda$, we obtain the 
upper limits on the coupling strengths, which are 
$\vert C_{bB}\vert < \sqrt {8\pi}$, 
$\vert C_{qW}\vert < \sqrt {4\pi}$, and  
$\vert C_{qB}\vert < \sqrt {4\pi}$.
They imply that to give the minimal contribution required by 
the $A_b$ and $R_b$ data the new physics scale cannot be too high. 
The  $1\sigma$ ($2\sigma$) upper bounds are found to be
\begin{eqnarray}
 \Lambda & \lsim &  1.4 (2.0){\rm TeV}~~({\rm for}~ O_{bB}),\\
 \Lambda & \lsim &  2.6 (5.9){\rm TeV}~~({\rm for}~ O_{qW}),
\end{eqnarray}
and 
\begin{eqnarray}
\Lambda & \lsim &  2.8 (6.2){\rm TeV}~~({\rm for}~ O_{qB}).
\end{eqnarray}
The upper bounds on the new physics scales for the three 
momentum-independent operators are much weaker; e.g.,
$\Lambda \lsim 10~(14)$ TeV at the $1\sigma$ ($2\sigma$) level for 
the operator $O_{\Phi b}$.      

\section{Correlated effects at LEP II, the NLC and the Tevatron}
\label{sec4}

The anomalous couplings in Eqs. (\ref{Vqq}) and (\ref{Wtb}) also 
contribute to the 
production cross sections for $e^+e^-\to Z^*,\gamma^* \to q\bar q$ ($q=b,t$) 
and $u\bar d\to W^* \to t \bar b$ as follows: 
\begin{eqnarray}\label{sigma_qq}
\delta\sigma(e^+e^-\to q\bar q)&=&\frac{3}{2}\beta_q\left \{
D_{\gamma\gamma}e_e^2 \left[ (3-\beta_q^2)e_q
                  (\delta g^{\gamma}_L+\delta g^{\gamma}_R)
                                        \right ]\right.\nonumber\\
& & 
   +D_{ZZ}(v_e^2+a_e^2)\left[(3-\beta_q^2)v_q(\delta g^Z_L+\delta g^Z_R)
   +2\beta_q^2 a_q (\delta g^Z_L-\delta g^Z_R)   \right ]\nonumber\\
& &
+D_{Z\gamma}e_e v_e\left [\frac{3-\beta_q^2}{2}
            \left (e_q (\delta g^Z_L+\delta g^Z_R)
 +v_q (\delta g^{\gamma}_L+\delta g^{\gamma}_R)\right )\right.\nonumber\\
& & \left. \left.
+\beta_q^2 a_q (\delta g^{\gamma}_L-\delta g^{\gamma}_R)
  \right ]\right \},
\end{eqnarray}
and
\begin{eqnarray}\label{sigma_tb}
\delta \hat{\sigma}(u\bar d\to t \bar b)&=&
 \frac{g^4}{384\pi}\frac{(\hat{s}-m^2_t)^2}
{\hat{s}^2(\hat{s}-m^2_W)^2}
\left [ 2(2\hat{s}+m^2_t)\delta g^W_L \right ].
\end{eqnarray}
Here $s$ and $\hat s$ are the squared center-of-mass energies
for $e^+e^-\to q\bar q$ and $u\bar d\to t \bar b$, respectively.
$v_f$ and $a_f$ represent, respectively, the vector and axial-vector 
$Zf\bar f$ couplings in the SM; i.e., 
$v_f\equiv (g^Z_L+g^Z_R)/2$ and $a_f\equiv (g^Z_L-g^Z_R)/2$.  
In Eq.(\ref{sigma_qq}) we have defined
\begin{eqnarray}
\beta_q&=&\sqrt{1-4m_q^2/s},\\
D_{\gamma\gamma}&=&\frac{4\pi \alpha^2(s)}{3s},\\
D_{ZZ}&=&\frac{G_F^2}{96\pi}\frac{sm_Z^4}{(s-m_Z^2)^2+(s\Gamma_Z/m_Z)^2},
\end{eqnarray}
and 
\begin{eqnarray}
D_{Z\gamma}&=&\frac{G_F\alpha(s)}{3\sqrt 2}\frac{m_Z^2(s-m_Z^2)}
              {(s-m_Z^2)^2+(s\Gamma_Z/m_Z)^2}.
\end{eqnarray}
The anomalous contribution to  
$R_b\equiv\sigma(e^+e^-\to b\bar b)/\sigma(e^+e^-\to q\bar q)$
and the forward-backward asymmetry $A^b_{\rm FB}$
at LEP II are given by 
\begin{eqnarray}
\frac{\delta R_b }{R_b^{\rm SM}}=(1-R_b^{\rm SM})
\frac{\delta \sigma(e^+e^-\to b\bar b)}{\sigma^{\rm SM}(e^+e^-\to b\bar b)},
\end{eqnarray}
and
\begin{eqnarray}
\frac{\delta A^b_{\rm FB}}{A^{\rm b, SM}_{\rm FB}}&=&
\frac{D_{Z\gamma}e_e a_e (e_b \delta g^Z_L-e_b\delta g^Z_R
 +g_L^Z \delta g^{\gamma}_L-g_R^Z\delta g^{\gamma}_R )
 +4D_{ZZ}v_e a_e (g_L^Z \delta g^Z_L-g_R^Z\delta g^Z_R)}
       {2D_{Z\gamma}e_e a_e e_b a_b+8D_{ZZ}a_e v_e v_b a_b}\nonumber\\
& & -\frac{\delta \sigma(e^+e^-\to b\bar b)}
      {\sigma^{\rm SM}(e^+e^-\to b\bar b)}.
\end{eqnarray}
The total hadronic cross section for $p\bar{p} \to t\bar{b} +X$
is evaluated by the convolution of the parton cross section and 
parton distribution functions.  Here we use the CTEQ3L parton distribution
functions~\cite{cteq3} with $\mu=\sqrt {\hat s}$.  The top quark mass is
taken to be 175 GeV.  The values of the other parameters are taken to be
$m_Z=91.187$, $m_W=80.33$, $G_F=1.16639\times 10^{-5}{\rm GeV}^{-2}$
and $\alpha=1/128$. 

Since the existence of both new right- and left-handed operators is
necessary to explain the LEP I data, we again assume only one pair of the
operators given in Eqs. (\ref{bB}) - (\ref{phiq3}) contributes at a time
in obtaining the limits on their contributions.
Subject to the limits derived from the data on $A_b$ and $R_b$, their
effect on $R_b$ and $A^b_{\rm FB}$ at LEP II, the $t\bar t$ production 
rate at the NLC, and the single top production cross section
$\sigma(p\bar p \to t\bar b+X)$  at the Tevatron can be 
evaluated. These effects are summarized in Table \ref{tab3}.

A few remarks regarding to the results in Table \ref{tab3} are appropriate:
\begin{description}
\item[{\rm(a)}] The right-handed operator $O_{bB}$ has large effects on 
                $R_b$ and $A^b_{\rm FB}$ at LEP II due to two reasons: 
                One is that it contributes to the $\gamma^* b_R\bar b_R$ 
                coupling and the $\gamma^*$ intermediate state gives
                the dominant contribution to $\sigma(e^+e^-\to b\bar b)$
                at LEP II.  The other is that the effects of $O_{bB}$ are
                enhanced by a factor $s/m_Z^2$, which is 4.3 for 
                $\sqrt s=189$ GeV at LEP II, due to its momentum dependent
                couplings.
The preliminary LEP II data at $\sqrt s=189$ GeV give \cite{opal}
\begin{eqnarray}\label{eq22}
& & R_b^{\rm exp}
   =0.167\pm 0.011({\rm stat.})\pm 0.008 ({\rm syst.}),~~
R_b^{\rm SM} =0.162,\\ \label{eq23}
& &  A^{\rm b,exp}_{\rm FB}=0.68\pm 0.21({\rm stat.})\pm 0.04 ({\rm syst.}),~~
A^{\rm b, SM}_{\rm FB}=0.56.
\end{eqnarray}
Note the large statistical error in $A^{b,exp}_{FB}$.  At the $1\sigma$ 
($2\sigma$) level the new physics effects are limited to the following
ranges 
\begin{eqnarray}\label{eq24}
-5\%(-14\%) \lsim \frac{\delta R_b}{R_b^{\rm SM}} \lsim 11\%~ (20\%),\\
\label{eq25}
-17\%(-55\%)\lsim \frac{\delta A^{\rm b}_{\rm FB}}{A^{\rm b, SM}_{\rm FB}}
                   \lsim 60\%~ (98\%).
\end{eqnarray}
From Table \ref{tab3} we see that the minimum contributions of $O_{bB}$,
 when it is used together with any one of the left-handed operators,  
lie outside the allowed ranges in Eqs. (\ref{eq24}) and (\ref{eq25}) 
at the $1\sigma$ level.  Hence the right-handed operator $O_{bB}$ is 
disfavored unless the central values of both $R^{exp}_b$ and $A^{b,exp}_{FB}$ 
are modified in the refined LEP II data analysis. 
The operator $O_{\Phi b}$, which is neither momentum dependent nor  
contributes to the $\gamma^* b_R\bar b_R$ coupling, 
makes much smaller contributions to $R_b$ and 
$A^b_{FB}$ at LEP II. As shown in  Table \ref{tab3},
when $O_{\Phi b}$ is used together with a left-handed operator
the deviation from the SM is allowed by the preliminary LEP II data.

\item[{\rm(b)}] The left-handed operator, $O_{qW}$, 
                when used to fit the anomalous $Zb_L\bar b_L$ coupling at the
                $Z$-pole, can give rise to a sizable effect in
                the cross section for $p\bar{p} \to t\bar{b} +X$ at
                the Fermilab Tevatron because its effect on $W^*t\bar b$ is  
                 enhanced by a 
                 factor $\hat s/m_Z^2$ ($\sqrt{\hat s}$ is the center-of-mass 
                energy of the parton-level process $u\bar{d} \to t\bar{b}$).
                Moreover, both $O_{qW}$ and $O_{qB}$
                can cause large effects in the production rate of $t\bar t$ 
                pairs at the 
                NLC because their anomalous $\gamma^* t_L\bar t_L$ coupling
                is enhanced by a factor $s/m_Z^2\approx 30$ 
                for $\sqrt s=500$ GeV at the NLC.   Note that the effect of
                $O_{qB}$ is about a factor two larger than that of $O_{qW}$. 

\item[{\rm(c)}] Due to the clean environment and anticipated large number of 
                the top pair events at the NLC it is possible to measure the 
                top 
                pair production rate at the level of a few percent\cite{NLC}.
              So the  $t\bar t$ production rate decrease caused by  
               $O_{qW}$ ($O_{qB}$),
                with the minimum $2\sigma$ limit of 3.7\% (7.2\%),  
                should be observable at the $2\sigma$ level. 
                In contrast, due to the backgrounds at hadron
                colliders, it will be challenging to measure the single top 
                production rate at the level of a few percent\cite{single}
                at the Tevatron. The decrease of 
                the single top production rate caused by  $O_{qW}$,
                with the minimum  $1\sigma$($2\sigma$) limit of 15\% (3\%),  
                should be observable at the $1\sigma$ level, but probably 
                not at the $2\sigma$ level.
                A detailed Monte Carlo analysis, with the consideration
                of all possible backgrounds, showed \cite{hikasa} 
                that the effects of  $O_{qW}$ on the single top production 
                rate are observable at Run 3, or Run 2b, 
                (30 fb$^{-1}$ luminosity) 
                at the $2\sigma$ level for 
                $C_{qW}/(\Lambda/{\rm TeV})^2 \gsim 0.5$. Consequently, the
                effects of  $O_{qW}$, with coupling strength in 
                the $2\sigma$ range, 
                $1.73 \gsim C_{qW}/(\Lambda/{\rm TeV})^2 \gsim 0.1$ in 
                 Eq.(\ref{limit2}),  will only marginally be 
                 observable in Run 3,  or Run 2b, at the Tevatron.

\item[{\rm(d)}] Among the left-handed operators 
                 the two operators that involve the Higgs field, 
                  $O_{\Phi q}^{(1)}$ and 
                 $O_{\Phi q}^{(3)}$, together with the right-handed 
                 operator, $O_{\Phi b}$, which also  involve the Higgs field,
                 indeed, do provide the required 
                 contribution to $Zb_L\bar b_L$ coupling but cause only
                 small effects at LEP II, the NLC, and the upgraded Tevatron;
                 only a few percent deviation from the SM in the $b$ 
                 forward-backward
                 asymmetry at LEP II.
\end{description}

\section{Discussion and Conclusion}
\label{sec5}

We have examined in some detail the possibility of new physics, characterized
by higher dimension operators, if the current data on the
$Zb\bar{b}$ couplings at the $Z$-pole are taken literally.  
The dimension six operators we considered consist of two right-handed
operators and four left-handed ones.  Both the right- and
left-handed operators are needed to explain the data.
We also examined the effects of the eight pairs of operator at high energy
colliders, LEP II, NLC, and upgraded Tevatron.  Since the operators have
sufficiently different behaviors at these higher energy colliders, 
their effects can mostly be distinguished, 
as can be seen from Table \ref{tab3}.  In particular,
the preliminary LEP II data disfavor four pairs of these operators, 
all of which involve the right-handed operator $O_{bB}$.

The four pairs which contain the operator $O_{\Phi b}$ also 
behave differently.   The pair
$O_{\Phi b}$ and $O_{qW}$ has only small effects on $\delta R_b$ and
$\delta A^b_{FB}$ but observable effects on $t\bar{t}$ production
at the NLC and single top production at the upgraded Tevatron.  The pair
$O_{\Phi b}$ and $O_{qB}$ has a strong effect on $t\bar{t}$ production
at the NLC, and an observable effect on $\delta A^b_{FB}$,
but a negligible effect on
$\delta R_b$ and no observable effects at the Tevatron.

The remaining two pairs, $O_{\Phi b}$ and $O^{(1)}_{\Phi q}$ and
$O_{\Phi b}$ and $O^{(3)}_{\Phi q}$, have very little effect on the high
energy quantities under consideration, except for a few percent deviation 
from the SM in
$A^b_{FB}$.  However, since these operators all contain the Higgs
field and, therefore, are related to the symmetry breaking sector, they
may well have something to do with new physics; at least,
our analysis indicates they are the most likely suspects.
The $Ht\bar{t}$ production at the NLC, 
$e^+e^-\to t\bar t H$, or $e^+e^-\to b\bar b H$, are useful processes
for detecting their existence, as these operators can induce
anomalous four-point couplings: $Z t\bar t H$ and $Z b\bar b H$. However,
the production rates for such processes depend on the Higgs mass,
which awaits discover.

In our analyses we have considered one pair of right- and left-handed
operators at a time.  If the two right-handed operators $O_{bB}$ and
$O_{\Phi b}$ are considered together $C_{bB}$ will be strongly
constrained by the LEP II data for the reasons discussed in remark (a)
of the previous section.  Then the effects of  $O_{bB}$ on the LEP I 
observables will be
suppressed relative to these of $O_{\Phi b}$.  Thus $O_{\Phi b}$ will 
remain to explain the LEP I data and our previous conclusion that
$O_{\Phi b}$ is the favored candidate for the right-handed new physics 
operator in  the
$Zb\bar b$ couplings is not changed.  Similarly, if we consider the four
left-handed operators together
we can obtain a bound on a linear combination of their coupling coefficients
from the LEP I data.  From top pair production at the NLC we can distinguish
$O_{qW}$ and $O_{qB}$ from the other two. From single top production at
the upgraded Tevatron we might also be able to single out $O_{qW}$. 
However, it will be difficult to distinguish $O_{\Phi q}^{(1)}$ from
$O_{\Phi q}^{(3)}$. 

In conclusion, if new physics effects exist in the $Zb\bar b$ vertex, as
indicated by the current measurements at the $Z$-pole, then: 
(1) The responsible right-handed operator could only be $O_{\Phi b}$, 
to be consistent with the LEP II data.
(2) The  responsible left-handed operators can be distinguished by 
measuring $t\bar{t}$ production rate at the NLC and single top production
at the Tevatron. (3) It is most likely that any new physics in $Zb\bar b$
couplings will involve the Higgs sector.
 
\section*{Acknowledgments}

JMY thanks Ken-ichi Hikasa for useful discussions. The work was supported in 
part by the Grant-in-Aid for Scientific Research (No.~10640243) and 
the Grant-in-Aid for JSPS Fellows (No.~97317) from the Japan Ministry of 
Education, Science, Sports, and Culture, and by the U.S.
Department of Energy, Division of High Energy Physics, under Grant
Nos. DE-FG02-91-ER4086, DE-FG02-94ER40817, and DE-FG02-92ER40730.  BLY
acknowledges the support by a NATO grant and the support and hospitality
provided to him by Dr. T.K. Lee at the National Center for Theoretical
Science, Hsin Chu, where part of the work has been done.  


\newpage
\begin{table}
\caption{The anomalous couplings induced by right-handed operators.
          $k$ is the momentum of the corresponding vector boson.}
\label{tab1}
\vspace{7mm}
\begin{center}
\begin{tabular}{l|cc} ~  & $ O_{bB}$ &  $O_{\Phi b} $ \\ \hline
    $ \delta g^Z_R(b)  $  
   &  $~~~\frac{4s_W^2c_W}{e}\frac{k^2}{\Lambda^2}C_{bB}~~~ $ 
   &  $~~~-\frac{4s_Wc_W}{e}\frac{vm_Z}{\Lambda^2}C_{\Phi b}~~~ $\\
 $\delta g^{\gamma}_R(b) $& $ -\frac{c_W}{e}\frac{k^2}{\Lambda^2}C_{bB} $& 0\\
 $\delta g^{\gamma}_R(t) $& 0 & 0\\
 $\delta g^{\gamma}_R(t) $& 0 & 0
\end{tabular}
\end{center}
\end{table}

\vspace{10mm}
\begin{table}
\caption{The anomalous couplings induced by left-handed operators.
         $k$ is the momentum of the corresponding vector boson.}
\label{tab2}
\vspace{7mm}
\begin{center}
\begin{tabular}{l|cccc} ~&$O_{qW}$&$ O_{qB} $&$ O_{\Phi q}^{(1)} $
&$ O_{\Phi q}^{(3)}$\\   \hline
   $ \delta g^Z_L(b)   $&$ ~\frac{2s_Wc_W^2}{e}\frac{k^2}{\Lambda^2}C_{qW}~ 
                  $&$ ~\frac{4s_W^2c_W}{e}\frac{k^2}{\Lambda^2}C_{qB}~
         $&$ ~-\frac{4s_Wc_W}{e}\frac{vm_Z}{\Lambda^2}C_{\Phi q}^{(1)}~
         $&$ ~-\frac{4s_Wc_W}{e}\frac{vm_Z}{\Lambda^2}C_{\Phi q}^{(3)}~$\\
$\delta g^{\gamma}_L(b)$&$  ~\frac{s_W}{2e}\frac{k^2}{\Lambda^2}C_{qW}~ 
                  $&$ ~-\frac{c_W}{e}\frac{k^2}{\Lambda^2}C_{qB}~$&$0$&$0$\\ 
$\delta g^Z_L(t)$&$  ~-\frac{2s_Wc_W^2}{e}\frac{k^2}{\Lambda^2}C_{qW}~ 
                  $&$ ~\frac{4s_W^2c_W}{e}\frac{k^2}{\Lambda^2}C_{qB}~
         $&$ ~-\frac{4s_Wc_W}{e}\frac{vm_Z}{\Lambda^2}C_{\Phi q}^{(1)}~
         $&$ ~\frac{4s_Wc_W}{e}\frac{vm_Z}{\Lambda^2}C_{\Phi q}^{(3)}~$\\
$\delta g^{\gamma}_L(t)$&$  ~-\frac{s_W}{2e}\frac{k^2}{\Lambda^2}C_{qW}~ 
                  $&$ ~-\frac{c_W}{e}\frac{k^2}{\Lambda^2}C_{qB}~$&$0$&$0$ \\
$\delta g^W_L   $&$ ~-\frac{s_W}{e}\frac{k^2}{\Lambda^2}C_{qW}~ 
                  $&$ 0 $& 0
          & $~\frac{v^2}{\Lambda^2}C_{\Phi q}^{(3)}~$
\end{tabular}
\end{center}
\end{table}

\newpage
\begin{table}
\caption{The ranges of correlated effects which are 
required by the data on $A_b$ and $R_b$ at the $Z$-pole. 
No contributions are indicated by `$-$'.}
\label{tab3}
\vspace{7mm}
\begin{center}
\begin{tabular}{c|rcccc} 
 &  &\multicolumn{2}{c} {LEP II ($189$ GeV)} 
  &NLC  ($500$ GeV)
  &Tevatron  ($2$ TeV) \\ 
&   &$\frac{\delta R_b}{R_b^{SM}}~(\%)$&
   $\frac{\delta A^b_{FB}}{A^{b,SM}_{FB}}~(\%)$&
   $\frac{\delta \sigma(e^+e^-\to t\bar t)}
                      {\sigma^{SM}(e^+e^-\to t\bar t)}~(\%)$ &
   $\frac{\delta \sigma(P\bar P\to t\bar b+X)}
                    {\sigma^{SM}(P\bar P\to t\bar b+X)}~(\%)$\\ \hline
  & & & & & \\
  $O_{bB}$, $O_{qW}$  & ~~$1\sigma$ & 
  $17\sim33$&$ -32\sim-65$&$ -19\sim-48$&$ -15\sim-39$ \\
 & $2\sigma$ &$ 
  9.4\sim41$&$-15\sim-81$&$ -3.7\sim-63$&$  -3.0\sim-51$  \\ \hline

 $O_{bB}$, $O_{qB}$  &  ~~$1\sigma$ &$ 
  21\sim44$&$ -36\sim-77$&$ -36\sim-93$&  $-$   \\
 & $2\sigma$ &$ 
  10\sim55 $&$ -16\sim-98 $&$  -7.2\sim-122 $& $-$  \\ \hline

 $O_{bB}$, $O_{\Phi q}^{(1)}$  & ~~$1\sigma$  &$ 
 21\sim42 $&$ -31\sim-64 $&$ 0.2\sim0.5$&  $-$   \\
 & $2\sigma$ &$ 
  10\sim52 $&$  -15\sim-81 $&$  0.04\sim0.7  $&  $-$   \\ \hline

 $O_{bB}$, $O_{\Phi q}^{(3)}$  &  ~~$1\sigma$  &$ 
  21\sim42$&$ -31\sim-64$&$-0.2\sim-0.5$&$  -1.0\sim-2.6$\\
 & $2\sigma$ & $
10\sim52 $&$  -15\sim-81 $&$ -0.04\sim-0.7 $&$ -0.2\sim-3.4$  \\ \hline

 $O_{\Phi b}$, $O_{qW}$  & ~~$1\sigma$   &$ 
  -3.2\sim-8.8  $&$  -2.7\sim-5.8$&$  -19\sim-48 $&$ -15\sim-39$\\
 & $2\sigma$ &$ 
   -0.4\sim-12 $&$  -1.2\sim-7.4 $&$  -3.7\sim-63 $&$  -3.0\sim-51$  \\ \hline

 $O_{\Phi b}$, $O_{qB}$ & ~~$1\sigma$  &$ 
0.8\sim1.7$&$  -7.5\sim-18$&$ -36\sim-93 $&  $-$     \\
 & $2\sigma$ &$ 
 0.4\sim2.2 $&$ -2.1\sim-24 $&$ -7.2\sim-122 $& $-$     \\ \hline

 $O_{\Phi b}$, $O_{\Phi q}^{(1)}$  & ~~$1\sigma$ &$ 
   0.04\sim-0.3$&$  -2.6\sim-5.5$&$ 0.2\sim0.5$&   $-$    \\
 & $2\sigma$ &$ 
   0.2\sim-0.5 $&$  -1.1\sim-6.9 $&$ 0.04\sim0.7 $&  $-$   \\ \hline

 $O_{\Phi b}$, $O_{\Phi q}^{(3)}$ &  ~~$1\sigma$ &$ 
   0.04\sim-0.3 $&$  -2.6\sim-5.5 $&$  -0.2\sim-0.5 $&$ -1.0\sim-2.6$ \\
 &$ 2\sigma $&$ 
   0.2\sim-0.5 $&$ -1.1\sim-6.9 $&$ -0.04\sim-0.7 $&$ -0.2\sim-3.4$  \\
\end{tabular}
\end{center}
\end{table}

\begin{figure}
\begin{center}
\psfig{figure=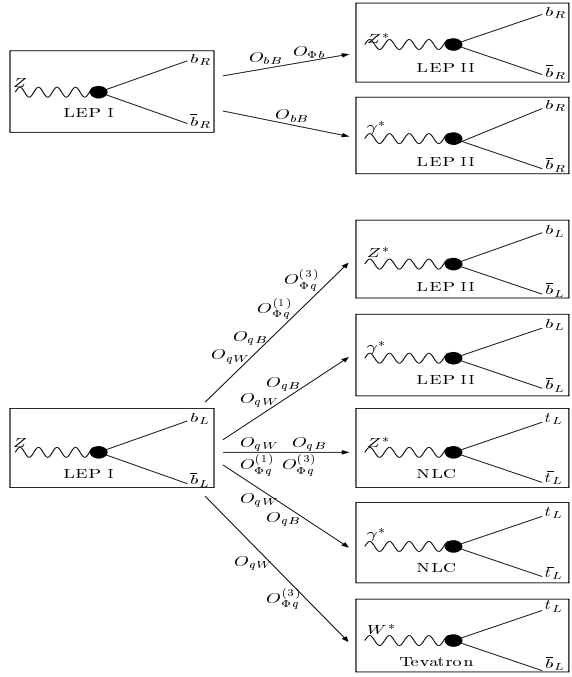,width=400pt,height=400pt,angle=0}
\end{center}
\vspace*{-2.5cm}
\caption{The diagrams showing the correlated effects of new physics
         in $Zb\bar b$ couplings.}
\vfil
\end{figure}
\end{document}